\documentclass[
reprint,
superscriptaddress,
amsmath,
amssymb,
aps,
prapplied,
]{revtex4-2}

\usepackage[T1]{fontenc}
\usepackage[utf8]{inputenc}

\usepackage{times}
\usepackage[unicode=true,breaklinks=true,colorlinks=true]{hyperref}
\hypersetup{citecolor=blue,urlcolor=blue}
\usepackage{amsmath}
\usepackage{amssymb}
\usepackage{gensymb}
\usepackage{graphicx}
\usepackage{bm}
\usepackage{bbold}
\usepackage{xcolor}

\usepackage{mathtools}
\usepackage[version=4]{mhchem}
\usepackage{tensor}
\usepackage{tabularx}
\newcommand{\overbar}[1]{\mkern 1.5mu\overline{\mkern-1.5mu#1\mkern-1.5mu}\mkern 1.5mu}

\begin{document}
	\title{Real-time adaptive sensing of nuclear spins by a single-spin quantum sensor}
	\author{Jingcheng Wang}
	\affiliation{School of Physics, Institute for Quantum Science and Engineering, International Joint Laboratory on Quantum Sensing and Quantum Metrology, Huazhong University of Science and Technology, Wuhan 430074, China}
	\affiliation{Hubei Key Laboratory of Gravitation and Quantum Physics, Huazhong University of Science and Technology, Wuhan 430074, China}
	\affiliation{Wuhan National High Magnetic Field Center, Huazhong University of Science and Technology, Wuhan 430074, China}
	\author{Dongxiao Li}
	\email{lidongxiao414@hust.edu.cn}
	\affiliation{School of Physics, Institute for Quantum Science and Engineering, International Joint Laboratory on Quantum Sensing and Quantum Metrology, Huazhong University of Science and Technology, Wuhan 430074, China}
	\affiliation{Hubei Key Laboratory of Gravitation and Quantum Physics, Huazhong University of Science and Technology, Wuhan 430074, China}
	\affiliation{College of physics science and technology, Shenyang Normal University, Shenyang 110034, China}
	\author{Ralf Betzholz}
	\affiliation{School of Physics, Institute for Quantum Science and Engineering, International Joint Laboratory on Quantum Sensing and Quantum Metrology, Huazhong University of Science and Technology, Wuhan 430074, China}
	\affiliation{Hubei Key Laboratory of Gravitation and Quantum Physics, Huazhong University of Science and Technology, Wuhan 430074, China}
	\author{Jianming Cai}
	\affiliation{School of Physics, Institute for Quantum Science and Engineering, International Joint Laboratory on Quantum Sensing and Quantum Metrology, Huazhong University of Science and Technology, Wuhan 430074, China}
	\affiliation{Hubei Key Laboratory of Gravitation and Quantum Physics, Huazhong University of Science and Technology, Wuhan 430074, China}
	\affiliation{Wuhan National High Magnetic Field Center, Huazhong University of Science and Technology, Wuhan 430074, China}
	\affiliation{ State Key Laboratory of Precision Spectroscopy, East China Normal University, Shanghai, 200062, China}
	\date{\today}
	
	\begin{abstract}
		Quantum sensing is considered to be one of the most promising subfields of quantum information to deliver practical quantum advantages in real-world applications. However, its impressive capabilities, including high sensitivity, are often hindered by the limited quantum resources available. 
		Here, we incorporate the expected information gain (EIG) and techniques such as accelerated computation into Bayesian experimental design (BED) in order to use quantum resources more efficiently. A simulated nitrogen-vacancy center in diamond is used to demonstrate real-time operation of the BED. Instead of heuristics, the EIG is used to choose optimal control parameters in real-time. Moreover, combining the BED with accelerated computation and asynchronous operations, we find that up to a tenfold speed-up in absolute time cost can be achieved in sensing multiple surrounding \ce{^{13}C} nuclear spins.
		Our work explores the possibilities of applying the EIG to BED-based quantum-sensing tasks and provides techniques useful to integrate BED into more generalized quantum sensing systems.
	\end{abstract}
	
	\maketitle
	
	\section{Introduction} 
	Quantum sensing, based on unique phenomena like quantum entanglement~\cite{Leibfried2004, Aasi2013}, quantum coherence~\cite{Huntemann2016} and quantum spin magnetometry~\cite{Taylor2008}, is used to measure physical quantities such as the electromagnetic~\cite{Taylor2008, Granata2016} and gravitational field \cite{Aasi2013}. 
	Compared to classical schemes, quantum sensing exhibits multiple advantages, such as a sensitivity exceeding the standard quantum limit~\cite{Leibfried2004, Aasi2013, Gefen2019, Bonato2016, Chu2020} and a microscopic spatial resolution~\cite{Taylor2008, Granata2016}.
	However, it is well known that the quantum resources quantum sensors rely on are limited due to unwanted effects~\cite{Degen2017}. This, in turn, has encouraged a plethora of efforts, such as the improvement of experimental apparatus~\cite{Vahlbruch2016}, the usage of dynamical decoupling~\cite{Du2009, Hirose2012, Zhao2012, Pham2012} and quantum heterodyne magnetometry \cite{Schmitt2017}, to enhance the robustness of the quantum sensors. Moreover, methods based on machine learning have recently gained increasing attention.
	
	Benefiting from specialized hardware~\cite{Jouppi2018}, ever-growing software support~\cite{Paszke2019}, and hyper-scale deep-learning models~\cite{Devlin2019}, machine learning is the building block of various recent breakthroughs in technology, such as artificial intelligence~\cite{Silver2016}, natural language processing~\cite{Wolf2020}, and computer vision~\cite{Bertinetto2016}. It is, therefore, not surprising that it also provides diverse applications in the field of quantum physics~\cite{Dunjko2016, Biamonte2017, Hush2017, Torlai2020} due to its high expressivity~\cite{pmlr-v70-raghu17a, Jordan2015}.
	In particular, as a remarkable example of machine learning, Bayesian experimental designs (BED) have been employed, among others, in processing the noisy readout of solid-state spins~\cite{Santagati2019} and reconstructing unknown quantum state of photonic systems~\cite{Yu2019}.
	
	On the other hand, the nitrogen-vacancy (NV) center spin in diamond, as an extremely well controllable ambient-condition quantum system, has been studied extensively in bio-imaging~\cite{Barnard2009, LeSage2013, Cai2013}, micro-scale magnetometry~\cite{Taylor2008, Schmitt2017}, and nuclear magnetic resonance~\cite{Staudacher2013,Jelezko2014,Bucher2020} due to its fluorescence brightness, good optical stability, high sensitivity to external magnetic fields, as well as low toxicity~\cite{Schmitt2017, Degen2017, Schirhagl2014}, and is likewise a suitable platform to demonstrate frameworks not limited to quantum sensing~\cite{Hou2019, Liu2019, Shi2010, Yang2020, Chu2021, Meinel2021, Gentile2021, Jiao2021}. 
	Notably, based on BED and NV centers, an increasing amount of frontier work has been put forward, such as static magnetic field magnetometry~\cite{Dushenko2020, Santagati2019}, quantum Hamiltonian learning~\cite{Wang2017, Joas2021}, and unknown quantum system learning~\cite{Gentile2021}. However, for more general quantum systems, conducting BED in real-time remains unexplored owing to the high demand of simulation resources and the lack of universally efficient heuristics. Heuristics that work well in one quantum sensing task could end up with sub-optimal results when employed in a similar, albeit different, setup. Besides, while it is well perceived that reducing the computational overhead in BED can free researchers from using more complex optimization process, the efforts spent in exploring this path are limited.
	
	In this work, we apply real-time BED on a simulated single NV-center spin interacting with a more generalized environment, such as multiple nearby nuclear spins and oscillating magnetic fields. To avoid the difficulty of finding suitable heuristics, we employ the expected information gain (EIG)~\cite{10.1214/aoms/1177728069,Bernardo1979,Chaloner1995,Ryan2015} as the utility function to obtain a system-agnostic guidance, which allows to optimize the experimental parameters efficiently. Moreover, the number of single-shot measurements required for the same uncertainty is reduced by a full order of magnitude, compared to non-adaptive protocols~\cite{Taminiau2012}. In addition, by virtue of a single consumer-grade general purpose graphical processing unit (GPGPU), the lengthy simulation time during real-time adaptive experiments can be substantially decreased in order to fulfill experimental requirements~\cite{Taminiau2012, Santagati2019}. To further reduce the time overhead induced by the adaptive design, we introduce asynchronous operation into the BED. Combining the accelerated computation and the asynchronous operation, we can translate the reduction in number of single-shot measurements into a speed-up in absolute time cost.
	
	The article is organized as follows. We begin by introducing our BED scheme including the EIG in Sec.~\ref{sec:BED}. In Sec.~\ref{ssec:nspins}, we then simulate some experiments to detect individual nuclear spins coupled to an NV center electron spin. We further test the protocol for the detection of oscillating magnetic fields using NV center as a magnetic sensor in Sec.~\ref{ssec:acmag}. In Sec.~\ref{sec:fastcalc}, the efforts in reducing the computational overhead are discussed. Finally, we summarize our work in Sec.~\ref{sec:sum}. 
	
	\section{Bayesian Experimental Design} 
	\label{sec:BED}
	
	Rigorous extensions of the concepts in BED have appeared
	in Refs.~\cite{Granade2012,Chaloner1995,Huan2013}. In our BED treatment, without loss of generality, we consider a two-level probe, initially prepared in the state $|0\rangle$, whose population in the basis states $\{|0\rangle$, $|1\rangle\}$ can be read~\cite{Degen2017}. The Hamiltonian of the system is denoted by $\hat{H}(\mathbf{x})$ with a set of parameters $\mathbf{x} = (x_1,\dots, x_n)$ which are being estimated~\cite{Oi2012}. 
	In order to estimate $\mathbf{x}$ accurately, repetitive measurements are required. In an experiment, these repetitions are treated as a series of events $E = (e_1,\dots, e_m)$. The control parameters and the read-out result of the $i$th single-shot event $e_i$ are written as $c_i$ and $d_i$, respectively. The control parameters $c_i$, which can, in principle, characterize any quantity that can be controlled and adjusted in the experiment, can be chosen from $\mathcal{C} = \{\text{c}_1,\dots, \text{c}_p\}$.
	
	According to the Bayes' theorem~\cite{sarkka_2013}, we can construct the probability distribution of $\mathbf{x}$ from a sequence of single-shot measurements, i.e.,
	\begin{equation}\label{PrD}
		\text{Pr}(\mathbf{x}|D) \propto \prod_{i=1}^{m} \text{Pr}(d_i|\mathbf{x}, c_i) \text{Pr}_0(\mathbf{x}),
	\end{equation}
	where $D = (d_1,\dots, d_m)$ represents sequential experiment data, $\text{Pr}_0(\mathbf{x})$ is the prior distribution of $\mathbf{x}$, and $\text{Pr}(d_i|\mathbf{x}, c_i)$ is the likelihood function~\cite{carlin2000bayes,mackay2003information} for our experiment. The probability that the state of the qubit probe of the experiment, denoted by $\vert\phi\rangle$, remains in the initial state $\vert0\rangle$ is given by $\text{Pr}(\vert\phi\rangle=\vert0\rangle)=\vert \langle 0 \vert \hat{U}(\mathbf{x},c_i) \vert 0 \rangle \vert^2$, where $\hat{U}(\mathbf{x},c_i)$ represents the time-evolution operator during the $i$th run of the experiment, where the evolution time itself is also included in the control parameter $c_i$ of the run. For an ideal read-out detection, the likelihood function is
	\begin{equation}
		\label{eq:ideal}
		\text{Pr}_\text{ideal}(d_i \vert \mathbf{x},c_i) = \vert \langle d_i\vert \hat{U}(\mathbf{x},c_i)\vert d_i\rangle\vert^2,
	\end{equation}
	where $d_i \in \{0, 1\}$ is the outcome of the $i$th single-shot measurement. However, inefficiency and noise in measurements are inevitable. Therefore, the likelihood function can be rewritten as
	\begin{gather}\label{eq:prnoise}
			\text{Pr}(0|\mathbf{x},c_i) =(p_0+p_1-1) \text{Pr}_\text{ideal}(0 \vert \mathbf{x},c_i)+(1-p_1), \\
				\text{Pr}(1|\mathbf{x},c_i) =1-\text{Pr}\left(0|\mathbf{x},c_i\right),
	\end{gather}
	where $p_0$ and $p_1$ represent the probabilities of a faithful detection of $|0\rangle$ and $|1\rangle$, viz., $p_0 = \text{Pr}(d=0 \vert \vert\phi\rangle=|0\rangle)$ and $p_1 = \text{Pr}(d=1 \vert \vert\phi\rangle=|1\rangle)$. 
	In order to improve the efficiency of our experiment, it is necessary to optimally choose the control parameter $c_i$ in terms of a utility function~\cite{Chaloner1995}. Explicitly, the EIG $\mathbb{E}_{\mathtt{Dist}_{\text{KL}}}$ is chosen as this utility function since it is able to achieve D-optimal designs~\cite{10.1214/aoms/1177728069} without finding specific heuristics for different models, even though the use of EIG does require efficient numerical realization to be suitable for non-trivial models~\cite{Ryan2015}. For an optimal $c_i$, it corresponds to the maxima of the EIG based on the previous measurements $D' = (d_1,\dots, d_{i-1})$, i.e., $c_i = \text{argmax}_{\text{c} \in \mathcal{C}} \mathbb{E}_{\mathtt{Dist}_{\text{KL}}}(\text{c}, D')$. In our setting, the EIG can be written as
	\begin{align}
		\label{eq:eig}
		&\mathbb{E}_{\mathtt{Dist}_{\text{KL}}}(\text{c}, D') = -\text{Pr}(1 \vert D',\text{c}) \int \text{Pr}(\mathbf{x}\vert D') \log[ \text{Pr}(1 \vert \mathbf{x}, \text{c})]  d\mathbf{x} \nonumber \\
		&-[1-\text{Pr}(1 \vert D', \text{c})]  \int\text{Pr}(\mathbf{x} \vert D') \log [1-\text{Pr}(1 \vert \mathbf{x}, \text{c})] d\mathbf{x},
	\end{align}
	where $\text{Pr}(1 \vert D',\text{c}) = \int \text{Pr}(1 \vert \mathbf{x}, \text{c}) \text{Pr}(\mathbf{x} \vert D')  d\mathbf{x}$ is the predicted probability~\cite{Granade2012} of detecting the state $|1\rangle$ based on the collected intermediate results.
	
	In summary, our real-time BED can be realized in the following steps: (i) Initially, we designate a suitable prior distribution $\text{Pr}_0(\mathbf{x})$ to conduct the iterative measurements. (ii) For each iteration $e_i$ in $E$, the optimal control parameter $c_i$ is found via the EIG. (iii) With this optimized control parameter, we carry out the measurement and obtain the result $d_i$, which can be further used to update the probability distribution $\text{Pr}\left(\mathbf{x}|D\right)$. (iv) Once our criteria are met, we stop the iteration and conclude the results. In detail, during this whole process, the numerical calculations necessary for the BED are realized by employing the sequential Monte Carlo (SMC) method~\cite{Ryan2015,doi:10.1080/01621459.1998.10473765} and the simulations of the experiments are performed using QuTiP~\cite{Johansson2013}. 
	
	\section{Applications of the BAYESIAN EXPERIMENTAL DESIGN}
	\label{sec:app}
	\subsection{NV center weakly coupled to nuclear spins}
	\label{ssec:nspins}
	
	\begin{figure*}[t]
		\centering
		\includegraphics[width=15cm]{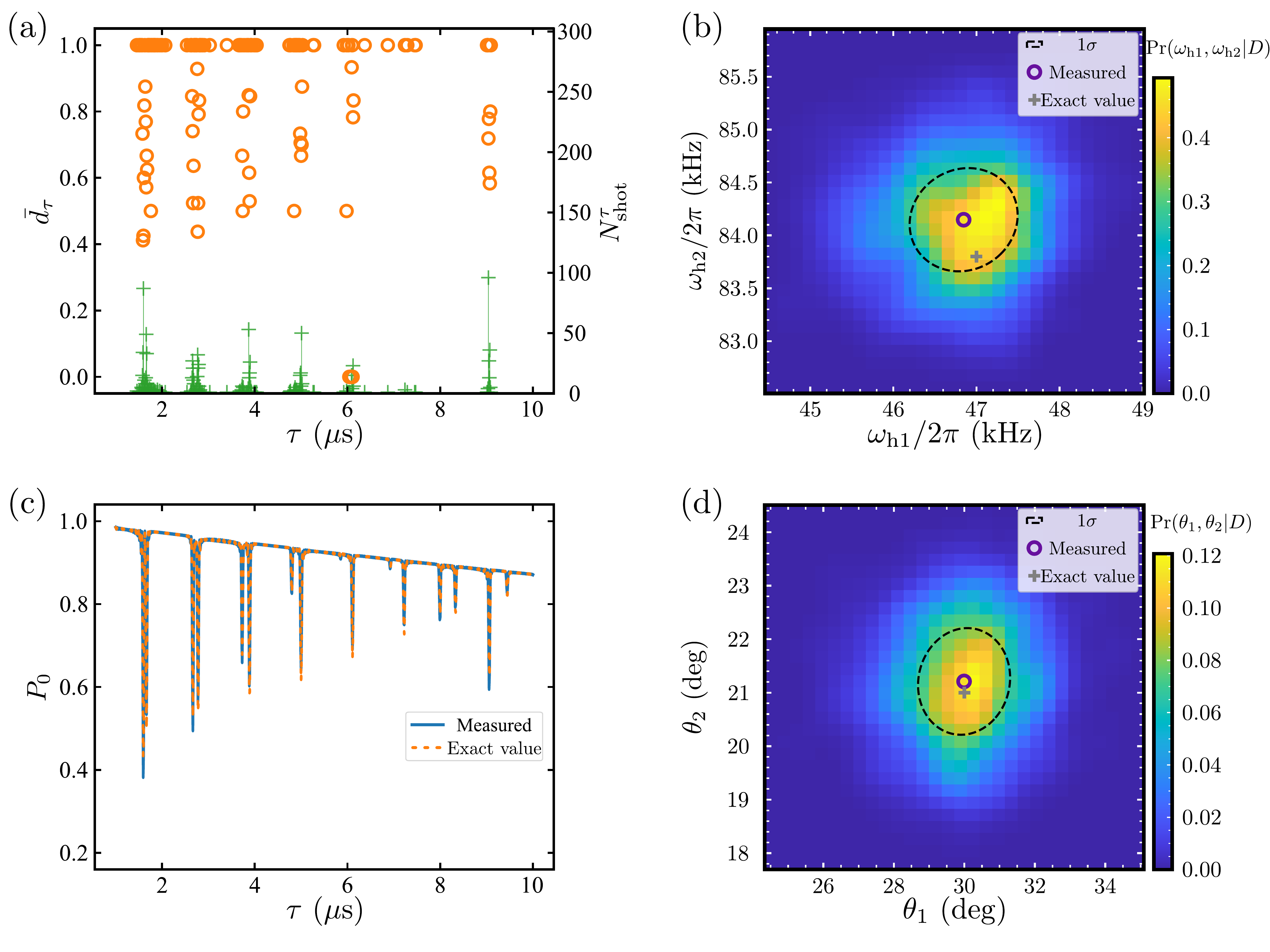}
		\caption{Simulation of the BED for sensing $n_\text{C}=2$ nuclear spins after $N_\text{ shot}=1050$ single-shot experiments are conducted. The exact values of parameters to be estimated are $\boldsymbol{\omega}_{\text{h}}/2\pi=(47.0,83.8)\,\text{kHz}$ and $\boldsymbol{\theta}= (30\degree, 21\degree)$. (a) Results of the single-shot measurements: the averaged measured results $\bar{d}_\tau$ versus $\tau$ (orange circles), a histogram of the number of measurements $N^{\tau}_\text{shot}$ for every $\tau$ (Green bars), and the number of measurements $N^{\tau}_\text{shot}$ per $\tau$ (green crosses). The dips of $\bar{d}_\tau$ correspond to the peaks of $N_{\text{shot}}^\tau$, showing that BED improves the efficiency by spending most resources on a subset of favorable control parameter values. (b) and (d) Histogram of the posterior distribution $\text{Pr}(\omega_{h1},\omega_{h2}|D)$ and $\text{Pr}(\theta_{1},\theta_{2}|D)$ in the form of marginal distributions based on the SMC method: Region with an approximately $68.27\%$ prediction confidence (black dashed ellipse), results of the estimation parameters reported by the BED (purple circle), and the exact values of the estimation parameters (gray cross). (c) Comparison between results of the BED and exact values in terms of the population of the probe: Estimated population $P_0$ reported by the BED (blue solid line) and exact values of the population (orange dashed line).}
		\label{fig:c13dist}
	\end{figure*}
	
	For the sake of providing evidence of the feasibility of our scheme, we consider the application to a specific physical system, namely a single NV-center spin in diamond as a sensing probe. By applying an external magnetic field, one can lift the degeneracy of the spin states $m_s = \pm 1$ of the ground-state manifold and encode the probe-qubit basis $\{|0\rangle$, $|1\rangle\}$ in the two sublevels $m_s=0, -1$. The entire system then consists of this probe and multiple \ce{^{13}C} nuclear spins~\cite{Taminiau2012} in the vicinity of the NV center. We consider the Larmor frequency of all these nuclear spins to be $\omega_\text{L}/2\pi=429.4\, \text{kHz}$ and neglect the weak interactions between different nuclear spins. Due to the oscillating nature of the signal produced by the hyperfine interactions, we adopt a quantum lock-in amplifier protocol~\cite{Kotler2011,Taminiau2012}. Furthermore, we assume that the coherence time of the probe qubit is extended to the value $T_2$ by employing an XY8-4 dynamical-decoupling sequence~\cite{AliAhmed2013}. To take the residual decoherence due to environmental effects into account, in our simulations we add a pure dephasing to the coherent dynamics. The resulting time evolution of the system density operator $\hat\rho$ is thereby described by  $\hat\rho(t)=\exp(\mathcal{L}t)\hat\rho(0)$, where the action of $\mathcal{L}$ is given by
	\begin{align}
		\label{eq:nspinlind}
		\mathcal{L}\hat\rho = &\frac{1}{i\hbar} \sum_{q=1}^{n_\text{C}}[\vert 0\rangle\langle 0\vert\hat{H}_{0q}+\vert1\rangle\langle 1\vert\hat{H}_{1q} , \hat\rho] \nonumber \\
		&+\frac{1}{4T_2} (2\hat{\sigma}_{z} \hat\rho \hat{\sigma}_{z}^{\dagger} -\hat{\sigma}_{z}^{\dagger} \hat{\sigma}_{z} \hat\rho - \hat\rho \hat{\sigma}_{z}^{\dagger} \hat{\sigma}_{z}),
	\end{align}
	where a total of $n_\text{C}$ nuclear spins are taken into consideration. The Hamiltonians of the hyperfine interaction have the form 
	\begin{gather}
		\hat{H}_{0q} = \hbar\omega_\text{L} \hat{\sigma}_{zq},\\
		\hat{H}_{1q}=\hbar[\omega_{\text{h}q}\cos(\theta_q)+\omega_\text{L}] \hat{\sigma}_{zq}+\hbar\omega_{\text{h}q}\sin(\theta_q) \hat{\sigma}_{xq},
	\end{gather}
	where $\hat{\sigma}_\kappa$ and $\hat{\sigma}_{\kappa q}$ $(\kappa=x,y,z)$ denote the Pauli operators for the qubit probe and the $q$th nuclear spin, respectively. In this system, the magnitudes $\boldsymbol{\omega}_{\text{h}} = (\omega_{\text{h}1},\dots, \omega_{\text{h}n_C})$ and the angles $\boldsymbol{\theta} = (\theta_{1}, \dots , \theta_{n_\text{C}})$ of the hyperfine interaction between the NV-center electron spin and the nuclear spins are the parameters to be estimated. This means for the present scenario we have $\mathbf{x}=(\boldsymbol{\omega}_{\text{h}},\boldsymbol{\theta})$. On the other hand, the control parameter of the experiment is $\tau$ which corresponds to the free evolution time $2\tau$ between the decoupling pulses (for details see Appendix~\ref{appendix:setup}).
	
	In terms of the incoherent time evolution generated by Eq.~\eqref{eq:nspinlind} and the dynamical-decoupling sequence XY8-4~\cite{AliAhmed2013} with a total evolution time of $64\tau$, when assuming instantaneous pulses, the expression $\vert \langle 0 \vert \hat{U}(\mathbf{x},c_i) \vert 0 \rangle \vert^2$ is replaced by $\langle 0\vert [\exp(64\mathcal{L}\tau_i)|0\rangle \langle 0|] \vert0\rangle$, where we keep in mind that the generator $\mathcal{L}$ depends on the parameters $(\boldsymbol{\omega}_{\text{h}},\boldsymbol{\theta})$. The likelihood function~\eqref{eq:prnoise} of a single-shot measurement event can thereby be derived and has the form 
	\begin{align}
		\label{eq:nspinpr}
		\text{Pr}&(0 \vert (\boldsymbol{\omega}_{\text{h}},\boldsymbol{\theta}),\tau_i)=\frac{p_0-2p_1+2}{3} \nonumber\\
		&+\frac{p_0+p_1-1}{2}  \left[ M((\boldsymbol{\omega}_{\text{h}}, \boldsymbol{\theta}),\tau_i)+\frac{1}{3} \right] e^{-(64\tau_i/T_2)},
	\end{align}
	and $\text{Pr}(1|(\boldsymbol{\omega}_{\text{h}}, \boldsymbol{\theta}),\tau_i)=1-\text{Pr}(0|(\boldsymbol{\omega}_{\text{h}}, \boldsymbol{\theta}),\tau_i)$. Here,we have defined
	\begin{align}
		M((\boldsymbol{\omega}_{\text{h}}, &\boldsymbol{\theta}),\tau_i)= \prod_{q=1}^{n_\text{C}} \left\{ 1 -2 \left[\frac{\sin(16\phi_{q})}{\cos\left(\frac{\phi_q}{2}\right)}\right]^2 \right.\nonumber\\
		& \left. \times \left[\frac{\omega_{\text{h}q} \sin(\theta _q) \sin\left(\frac{\tilde{\omega}_q \tau_i}{2}\right) \sin\left(\frac{\omega_\text{L} \tau_i}{2}\right)}{\tilde{\omega}_q  }\right]^2 \right\}
	\end{align}
	 where the phases $\phi_q$ are determined by
	\begin{align}
		\cos(\phi_q)=&\cos(\tilde{\omega}_q \tau_i) \cos(\omega_\text{L}\tau_i )\nonumber\\
		&-\frac{\omega_{\text{h}q} \cos(\theta_q)+\omega_\text{L}}{\tilde{\omega}_q} \sin(\tilde{\omega}_q \tau_i) \sin(\omega_\text{L} \tau_i)
	\end{align}
	and $\tilde{\omega}_q=[\omega_{\text{h}q}^{2}+\omega_\text{L}^{2}+2 \omega_{\text{h}q} \omega_\text{L} \cos(\theta_q)]^{1/2}$~\cite{Taminiau2012}.
	
	According to the SMC method, a probability distribution can be represented by a set of $n_\text{p}$ particles~\cite{Andrieu2010, doi:10.1080/01621459.1998.10473765}. Utilizing such a method combined with Eq.~(\ref{PrD}), one obtains
	\begin{equation}
	\label{eq:SMC}
		\text{Pr}((\boldsymbol{\omega}_{\text{h}}, \boldsymbol{\theta}) \vert D) \approx \sum_{k=1}^{n_\text{p}} w_{k} \delta((\boldsymbol{\omega}_{\text{h}}, \boldsymbol{\theta})-(\boldsymbol{\omega}_{\text{h}k}, \boldsymbol{\theta}_k)),
	\end{equation}
	in which $w_{k}$, fulfilling $\sum_k^{n_\text{p}}w_k=1$, and $(\boldsymbol{\omega}_{\text{h}k}, \boldsymbol{\theta}_k)$ are the weight and location of the $k$th particle, respectively. Combined with Eq.~\eqref{eq:eig}, the EIG can thereby be written as 
	\begin{align}\label{EKL}
		\mathbb{E}_{\mathtt{Dist}_{\text{KL}}}(\tau, D')=&-\bigg[1-\sum_{k=1}^{n_\text{p}} w_{k} \text{Pr}(0 \vert (\boldsymbol{\omega}_{\text{h}k}, \boldsymbol{\theta}_k), \tau)\bigg] \nonumber \\
		&\times \sum_{k=1}^{n_\text{p}} w_{k} \log[1 -\text{Pr}(0 \vert (\boldsymbol{\omega}_{\text{h}k}, \boldsymbol{\theta}_k), \tau)] \nonumber \\
		&-\sum_{k=1}^{n_\text{p}} w_{k} \text{Pr}(0 \vert (\boldsymbol{\omega}_{\text{h}k}, \boldsymbol{\theta}_k), \tau) \nonumber \\
		&\times \sum_{k=1}^{n_\text{p}} w_{k} \log[\text{Pr}(0 \vert (\boldsymbol{\omega}_{\text{h}k}, \boldsymbol{\theta}_k), \tau)].
	\end{align}
	Furthermore, by virtue of Eq.~(\ref{EKL}) and Eq.~\eqref{PrD}, for the $i$th run of the experiment, we can filter the optimal control parameter $\tau_{i} = \text{argmax}_{\tau} \mathbb{E}_{\mathtt{Dist}_{\text{KL}}}(\tau, D')$ to accumulate measurement events $E$ and obtain the posterior distribution $\text{Pr}((\boldsymbol{\omega}_{\text{h}}, \boldsymbol{\theta})|D)$. Finally, the average of the parameters $(\boldsymbol{\omega}_{\text{h}}, \boldsymbol{\theta})$ can be calculated according to $\bar{\boldsymbol{\omega}}_{\text{h}} = \sum_{k=1}^{n_\text{p}} w_{k} \boldsymbol{\omega}_{\text{h}k}$ and $\bar{\boldsymbol{\theta}} = \sum_{k=1}^{n_\text{p}} w_{k} \boldsymbol{\theta}_{k}$. The corresponding uncertainties $\Delta\boldsymbol{\omega}_{\text{h}}=(\Delta\omega_{\text{h}1},\dots, \Delta\omega_{\text{h}n_\text{C}})$ and $\Delta\boldsymbol{\theta}=(\Delta\theta_1,\dots, \Delta\theta_{n_\text{C}})$ can be derived from the particles in the SMC method as
	\begin{gather}
		\Delta^2\omega_{\text{h}q}=\sum_{k} w_{k} (\omega_{\text{h}qk}-\bar{\omega}_{\text{h}q})^2, \\
		\Delta^2\theta_{q}=\sum_{k} w_{k} (\theta_{qk}-\bar{\theta}_{q})^2.
	\end{gather}
	
	For the sake of evaluating the performance of our BED, we first set the number of nuclear spins to $n_\text{C}=2$ as an example and simulate experiments based on these settings under an XY8-4 dynamical-decoupling sequence with a decoherence time $T_{2} = 3\,\text{ms}$. The control parameter $\tau$ takes values ranging from $1\,\mu\text{s}$ to $10\, \mu\text{s}$ with a step size $\Delta\tau = 10\, \text{ns}$, i.e., $\mathcal{C} = \{1, 1.01, 1.02,\dots, 10\}\,\mu\text{s}$, whereas the pulse amplitudes satisfy the condition $t_\text{pulse} \ll \tau$. To reduce the number of iterations before convergence, faithful read-out operations are used, viz., $p_0=p_1=1$. The EIG is updated once every 15 single-shot measurements to take full advantage of the computational resources. Additional details of the simulation setup are reported in Appendix~\ref{appendix:setup} and the values of the parameters to be estimated that we use are $\boldsymbol{\omega}_{\text{h}}/2\pi = (47.0, 83.8)$\,kHz and $\boldsymbol{\theta}=(30\degree, 21\degree)$ based on realistic parameters reported in Ref.~\cite{Taminiau2012}. 
	During this experiment, for each single-shot event $e_i$, we first conduct the measurement with $\tau_i$, record the result $d_i$, and then update $\text{Pr}((\boldsymbol{\omega}_{\text{h}}, \boldsymbol{\theta})|D)$ accordingly, where the choice of $\tau_i$ is optimized via $\mathbb{E}_{\mathtt{Dist}_{\text{KL}}}$. 
	After $N_{\text{shot}}=1050$ measurements, the number of measurements $N_{\text{shot}}^\tau$ for the control parameter $\tau$ in $\mathcal{C}$ are plotted in Fig.~\ref{fig:c13dist}(a) as crosses. 
	From this, we obtain the averaged results $\bar{d}_\tau$ (empty circles) by $\bar{d}_\tau= \sum_j^{N_{\text{shot}}^\tau} d_{\tau j} / N_{\text{shot}}^\tau $, where $d_{\tau j}$ is the $j$th  result of all single-shot measurements conducted at $\tau$. It can be seen that the dips of $\bar{d}_\tau$ correspond to the peaks of $N_{\text{shot}}^\tau$. This indicates that our BED improves the efficiency by spending most resources on measurements under an optimized subset of control-parameter values. 
	Furthermore, we plot the posterior distribution $\text{Pr}((\boldsymbol{\omega}_{\text{h}}, \boldsymbol{\theta})|D)$ after $N_{\text{shot}}=1050$ in Fig.~\ref{fig:c13dist}(b) and (d) in the form of marginal distributions. Based on $\text{Pr}((\boldsymbol{\omega}_{\text{h}}, \boldsymbol{\theta})|D)$, we obtain the final estimation results of $(\boldsymbol{\omega}_{\text{h}}, \boldsymbol{\theta})$ and show the population of the probe $P_0=\langle 0\vert [\exp(64\mathcal{L}\tau)|0\rangle \langle 0|] \vert0\rangle$, obtained under the dynamics generated by Eq.~\eqref{eq:nspinlind} in Fig.~\ref{fig:c13dist}(c) as a blue solid line. Moreover, we also depict the population $P_0$ derived by the theoretical values of $(\boldsymbol{\omega}_{\text{h}}, \boldsymbol{\theta})$ in Fig.~\ref{fig:c13dist}(c) as an orange dashed line. The coinciding behavior of the two curves demonstrates the validity of our scheme. 
	
\begin{figure}[t]
		\centering
		\includegraphics[width=\linewidth]{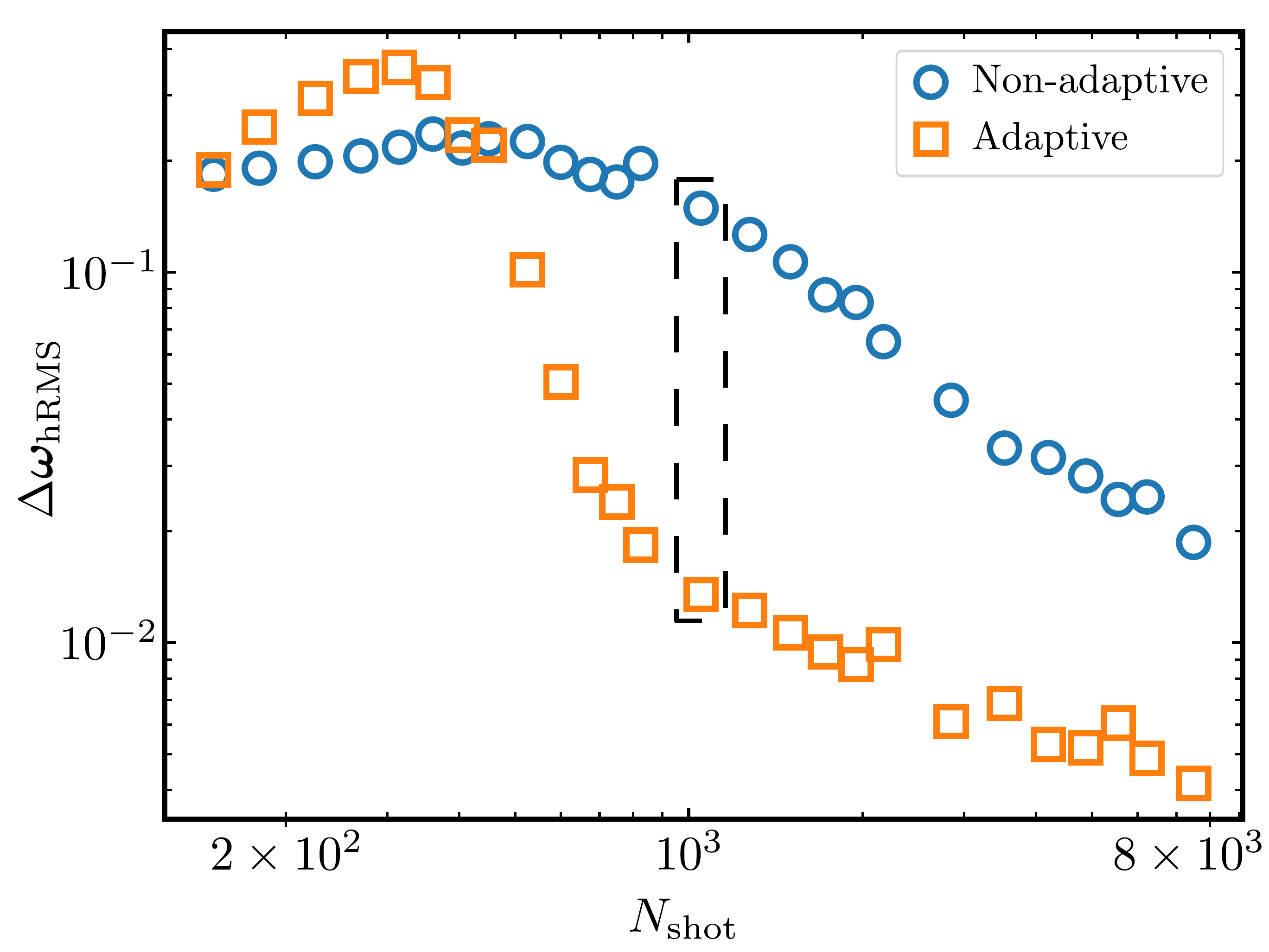}
		\caption{Comparison of the median of the averaged relative uncertainty $\Delta\boldsymbol{\omega}_\text{hRMS}$ between the BED and a non-adaptive protocol for sensing $n_\text{C}=2$ nuclear spins with 300 runs of the experiment. Orange squares: $\Delta\boldsymbol{\omega}_\text{hRMS}$ for the BED in dependence of the number of single-shot measurements $N_\text{shot}$. Blue empty circles: $\Delta\boldsymbol{\omega}_\text{hRMS}$ for a non-adaptive protocol versus $N_\text{shot}$. The exact values of $(\boldsymbol{\omega}_{\text{h}}, \boldsymbol{\theta})$ are the ones from Fig.~\ref{fig:c13dist}. At $N_\text{shot}=1050$ (highlighted by the dashed square), an averaged shot count per unique $\tau$ is $1.17$. The non-adaptive method fails to meet a relative uncertainty goal of $\Delta\boldsymbol{\omega}_\text{hRMS} \leq 10\%$, whereas the BED achieves $\Delta\boldsymbol{\omega}_\text{hRMS} < 2\%$.}
		\label{fig:c13scl}
	\end{figure}

	As a next step, in order to evaluate the relative accuracy of our BED, we repeat this experiment $300$ times. During the $j$th repetition, we record $\bar{\boldsymbol{\omega}}_{\text{h}j}$, $\bar{\boldsymbol{\theta}}_j$, $\Delta\boldsymbol{\omega}_{\text{h}j}$, and $\Delta\boldsymbol{\theta}_{j}$ against the number of single-shot measurements $N_{\text{shot}}$. Here, we use the mean relative uncertainties given by $\bar{\Delta}^2\omega_{\text{h}j}=\sum_{q=1}^{n_\text{C}} \Delta^2\omega_{\text{h}qj}/n_\text{C}$ and $\bar{\Delta}^2\theta_j=\sum_{q=1}^{n_\text{C}} \Delta^2\theta_{qj}/n_\text{C}$. After $300$ repetitions, we calculate the median $\Delta\boldsymbol{\omega}_\text{hRMS}$ of the recorded relative uncertainty $(\bar{\Delta}^2\omega_{\text{h}1},\dots,\bar{\Delta}^2\omega_{\text{h}300})$ and compare the results (empty squares) with those of the traditional non-adaptive protocol without a control parameter optimization (empty circles) in Fig.~\ref{fig:c13scl}. At $N_\text{shot}=1050$ (highlighted by a dashed square), an averaged shot count per unique $\tau$ is $1.17$, i.e., $\bar{N}_{\text{shot}}^{\tau}=N_\text{shot}/|\mathcal{C}|\approx1.17$. Here, we find that the non-adaptive method has failed to meet a relative uncertainty goal of $\Delta\boldsymbol{\omega}_\text{hRMS} \leq 10\%$ at such $\bar{N}_{\text{shot}}^{\tau}$. In contrast, our BED can achieve a relative uncertainty $\Delta\boldsymbol{\omega}_\text{hRMS} < 2\%$, which manifests that our BED is capable of utilizing the limited experimental resources efficiently.

\begin{figure}
		\centering
		\includegraphics[width=\linewidth]{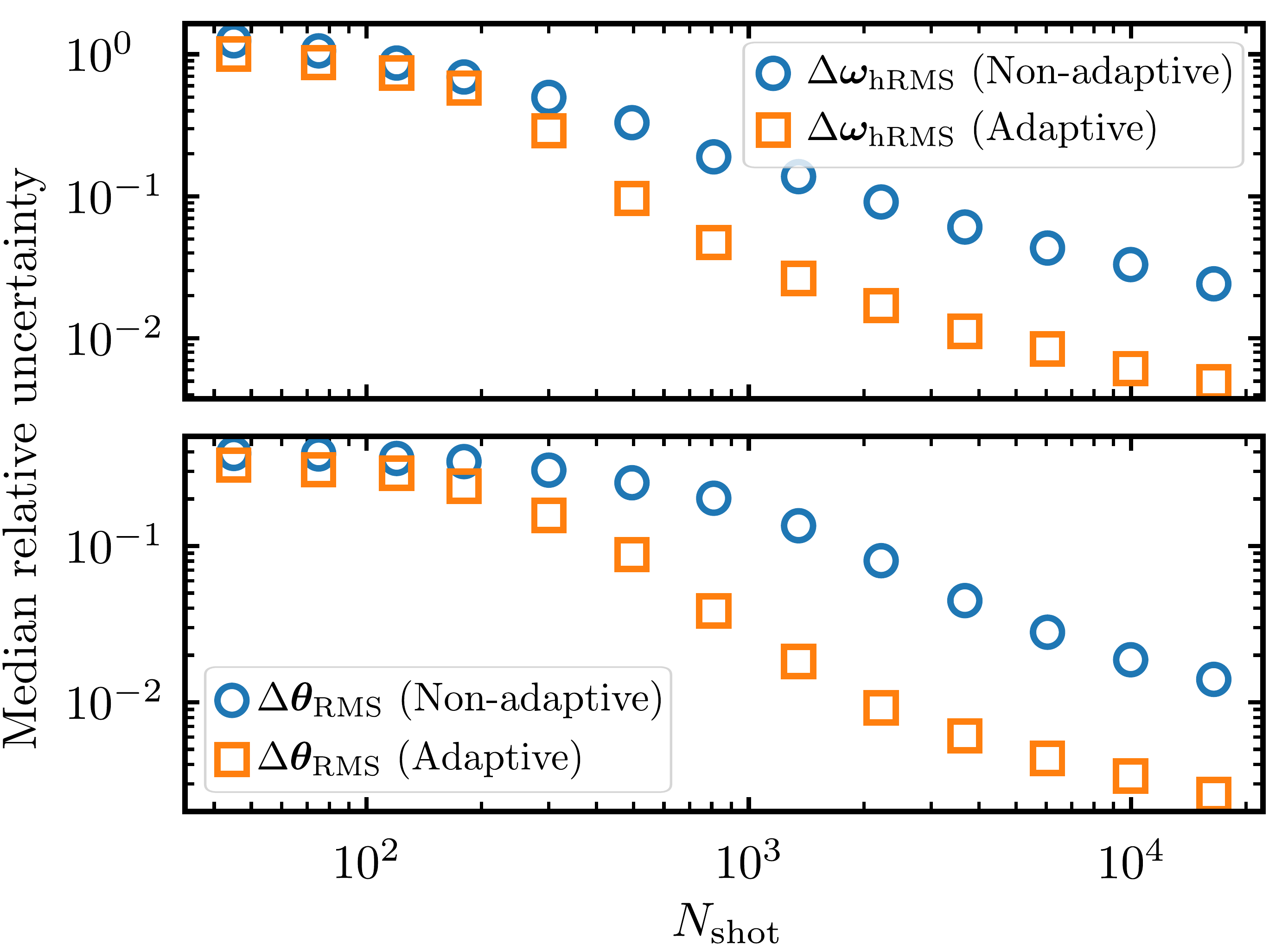}
		\caption{Comparison of the median of the averaged relative uncertainties $\Delta\boldsymbol{\omega}_\text{hRMS}$ and  $\boldsymbol{\theta}$ between the BED (orange squares) and a non-adaptive protocol (blue circles) for sensing $n_\text{C}=2$ nuclear spins based on $N_\text{bench}=900$ different combinations of $(\boldsymbol{\omega}_\text{h},\boldsymbol{\theta})$. Top: Dependence of $\Delta\boldsymbol{\omega}_\text{hRMS}$ on the number of single-shot measurements $N_\text{shot}$. Bottom: Dependence of $\Delta\boldsymbol{\theta}_\text{RMS}$ on $N_\text{shot}$. The set of exact parameters can be found at Appendix~\ref{appendix:setup}.}
		\label{fig:figc13rnd2rel}
	\end{figure}

	Furthermore, in order to quantify the performance of the BED in this model, we also randomly generate $N_{\text{bench}} = 900$ different combinations of $(\boldsymbol{\omega}_{\text{h}}, \boldsymbol{\theta})$ with $n_\text{C} = 2$. Instead of repeating experiments under the same $(\boldsymbol{\omega}_{\text{h}}, \boldsymbol{\theta})$ multiple times, we carry out one experiment for each combination of $(\boldsymbol{\omega}_{\text{h}}, \boldsymbol{\theta})$ generated beforehand. After conducting the experiments based on $N_{\text{bench}}$ combinations, we denote the median of the mean relative uncertainties $(\bar{\Delta}^2\omega_{\text{h}1},\dots,\bar{\Delta}^2\omega_{\text{h}N_{\text{bench}}})$ and $(\bar{\Delta}^2\theta_{1},\dots,\bar{\Delta}^2\theta_{N_{\text{bench}}})$ as $\Delta\boldsymbol{\omega}_\text{hRMS}$ and $\Delta\boldsymbol{\theta}_\text{RMS}$. We then plot $\Delta\boldsymbol{\omega}_\text{hRMS}$ and $\Delta\boldsymbol{\theta}_\text{RMS}$ of our BED versus $N_{\text{shot}}$ in Fig.~\ref{fig:figc13rnd2rel} as orange empty squares. For comparison, we simulate experiments of using a non-adaptive protocol under the same settings, and plot their scaling of the relative uncertainties in Fig.~\ref{fig:figc13rnd2rel} as blue empty circles.
	The general relative uncertainty $\Delta_\text{RMS}=\max\{\Delta\boldsymbol{\omega}_\text{hRMS},\Delta\boldsymbol{\theta}_\text{RMS}\}$ based on our method is $2.6\%$ while for the non-adaptive method we find $\Delta_\text{RMS} = 13.8\%$ at $N_\text{shot} = 1350$. Overall, by adopting our BED, a reduction of over $90\%$ in $N_\text{shot}$ for the same relative uncertainty can be observed.
	
\begin{figure}
		\centering
		\includegraphics[width=\linewidth]{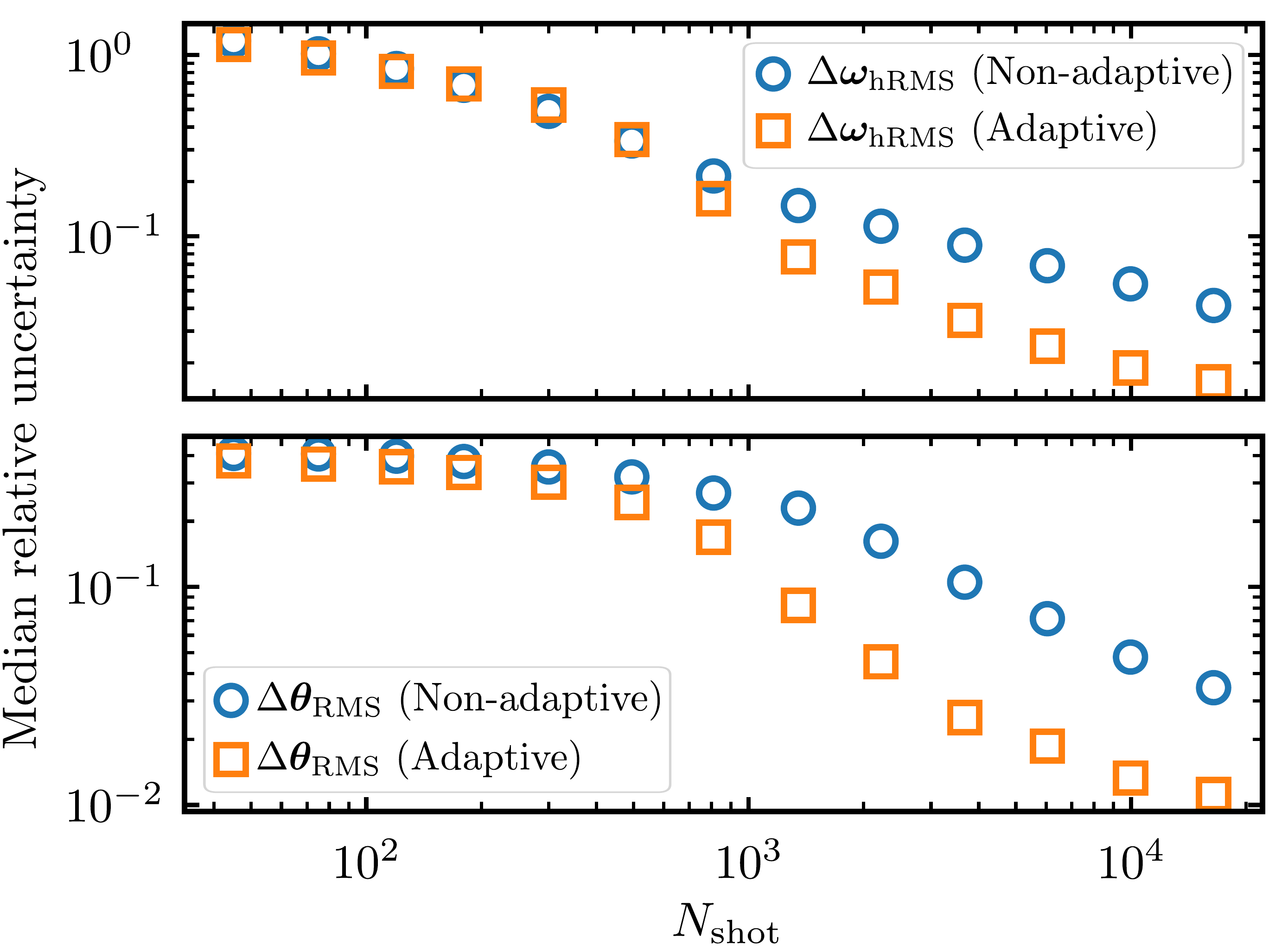}
		\caption{Comparison of the performance of the BED (orange squares) with a non-adaptive protocol (blue circles) for sensing $n_\text{C}=3$ nuclear spins based on $N_\text{bench}=1140$ different combination of $(\boldsymbol{\omega}_\text{h},\boldsymbol{\theta})$. Top: Median of the averaged relative uncertainty $\Delta\boldsymbol{\omega}_\text{hRMS}$ versus the number of single-shot measurements $N_\text{shot}$. Bottom: Median of the averaged relative uncertainty $\Delta\boldsymbol{\theta}_\text{RMS}$ versus $N_\text{shot}$.}
		\label{fig:figc13rnd3rel}
	\end{figure}

	To further investigate the abilities of our BED, we repeat the above test with $n_\text{C}=3$ neighboring nuclear spins and plot $\Delta\boldsymbol{\omega}_\text{hRMS}$ as well as $\Delta\boldsymbol{\theta}_\text{RMS}$ versus $N_\text{shot}$ in Fig.~\ref{fig:figc13rnd3rel}. For $\Delta\boldsymbol{\omega}_\text{hRMS}$, a reduction of $65\%$ is observed at $N_\text{shot} = 9975$, while a reduction of $72\%$ in $\Delta\boldsymbol{\theta}_\text{RMS}$ can be seen at the same shot count. This shows that our BED significantly improves the estimation accuracy of the parameters and further demonstrates the feasibility of our scheme.
	
	\subsection{Oscillating magnetic field detection with NV center}
	\label{ssec:acmag}
	
	Not only is our BED useful when an accurate likelihood function is available, but it also has advantages over traditional methods when combined with imperfect likelihood functions. In this subsection, we discuss our BED in a scenario where a qubit probe encoded in a single NV-center spin is used to sense an oscillating magnetic field.
	
	The probe is subject to an external oscillating magnetic field as well as additional sources of noise. In this case, the effective Hamiltonian of the probe after aligning the NV-center axis with the magnetic field~\cite{Jakobi2017} can be written as 
	\begin{equation}
		{\hat{H}}_{\text{eff}}(t) = \frac{\hbar\gamma B}{2}  h(t)  \cos( \omega t + \varphi) {\hat{\sigma}}_z + {\hat{H}}_{\text{noise}}(t),
	\end{equation}
	in which $\gamma/2\pi = 28.03 \, \text{MHz/mT}$ is the gyromagnetic ratio of the electron spin and $h(t)$ is the modulation function determined by the dynamical-decoupling pulses. In this case, the parameters that have to be estimated in order to determine the dynamics of the system are $\mathbf{x}=(B,\omega,\varphi)$, i.e., the magnitude, frequency, and initial phase of the magnetic field to be measured. However, since we assume that we do not record timestamps for each measurement, we have to average over the phase $\varphi$ (see Appendix~\ref{appendix:ac}). Effectively, the parameters to be estimated are thereby only $(\omega,B)$. Furthermore, ${\hat{H}}_{\text{noise}}(t)$ in the above Hamiltonian represents the effective action of noise on the probe
	and we apply an XY8 dynamical-decoupling pulse sequence~\cite{Kotler2011} in order to protect the probe from this unwanted environmental influence to some degree~\cite{Degen2017}. The dynamical decoupling sequences used here serve two purposes: amplifying the oscillating magnetic field being measured and extending the coherence time of the sensor. The time $\tau$ related to the free-evolution time of the pulse sequence, again, is the control parameter. As shown in Appendix~\ref{appendix:ac}, if we assume a coherence time $T_2$, the averaged ideal probability~\eqref{eq:ideal} of preserving the initial state $\left|0\right>$ reads
	\begin{equation}
		\label{eq:acprideal}
		\text{Pr}_\text{ideal}(0 \vert (\omega, B), \tau_i) = \frac{1}{2}[1+J_0(a_i)e^{-(16 \tau_i/T_2)}],
	\end{equation}
	with the Bessel function $J_0$ and where we have defined
		\begin{equation}\label{eq:phis}
		a_i=\frac{16B\gamma}{\omega}\sin^3\left(\frac{\omega\tau_i}{2}\right)
		\sum_{k=3,5,11,13} \cos\left(\frac{\omega\tau_i}{2}k\right).
	\end{equation}
	
\begin{figure}[t]
		\centering
		\includegraphics[width=\linewidth]{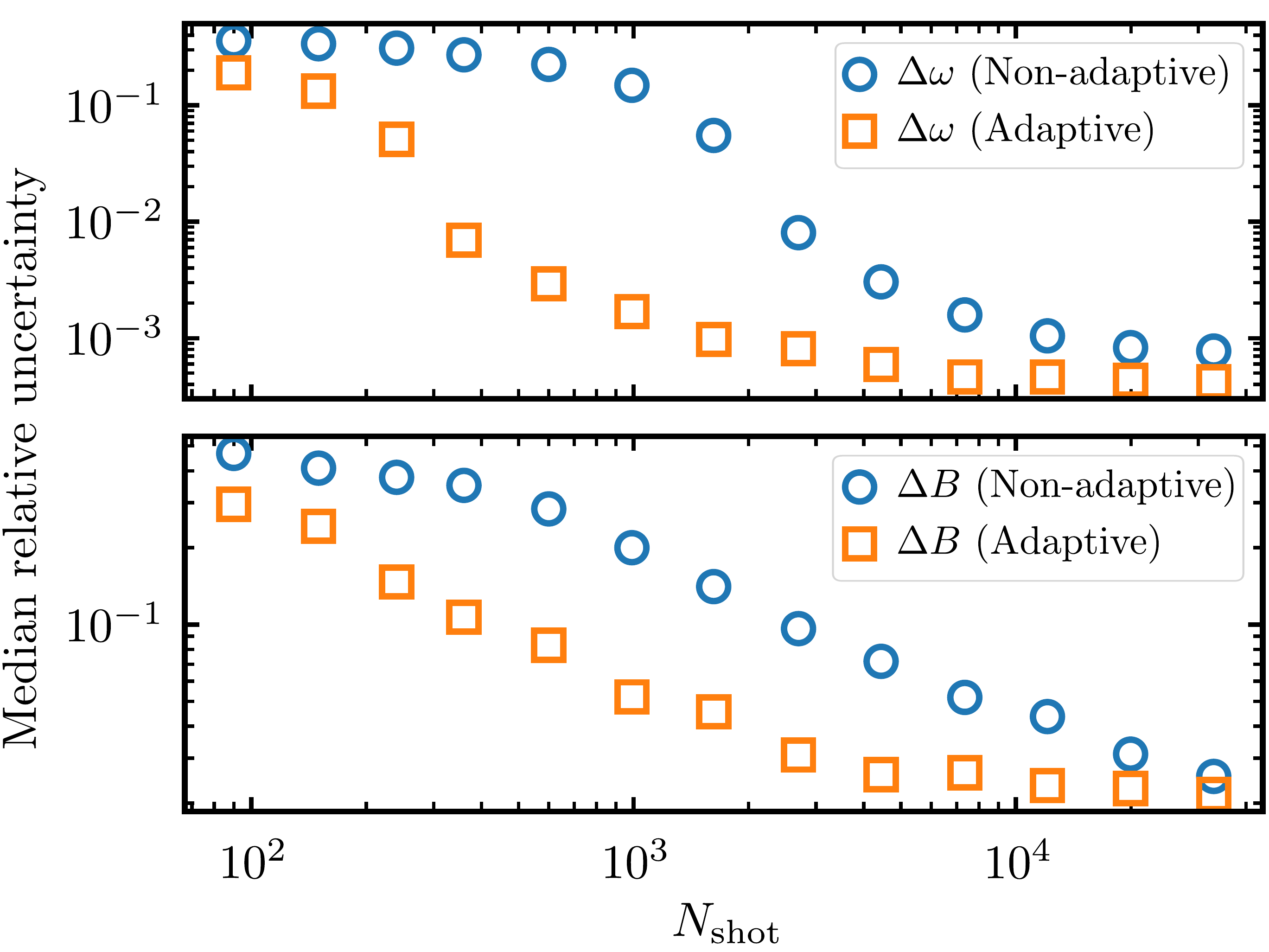}
		\caption{Comparison of the performance of the BED (orange squares) with a non-adaptive protocol (blue circles) for sensing an oscillating magnetic field with $N_\text{bench}=400$ different combinations of $(\omega, B)$. Top: Median of the relative uncertainty $\Delta\omega$ versus the number of single-shot measurements $N_\text{shot}$. Bottom: Median of the relative uncertainty $\Delta B$ versus $N_\text{shot}$.}
		\label{fig:acrel}
	\end{figure}

	However, to attain a suitable numerical efficiency in our implementation, we truncate the series expansion of the Bessel function at the sixth order. With the above ideal probability, the likelihood function~\eqref{eq:prnoise} then takes the form
	\begin{align}
	    &\text{Pr}(0 \vert (\omega, B),\tau_i)=\frac{p_0-p_1+1}{2} \nonumber\\
    	&+\frac{p_0+p_1-1}{2}  \left(1-\frac{1}{4}a_i^2+\frac{1}{64} a_i^{4}-\frac{1}{2304} a_i^{6}\right) e^{-(16\tau_i/T_2)}
	\end{align}
and $\text{Pr}(1 \vert (\omega, B), \tau_i)=1-\text{Pr}(0 \vert (\omega, B), \tau_i)$. Here, we mention that this is a faithful modeling of the system only when $N_{\text{shot}}^\tau \rightarrow \infty$ due to the average over $\varphi$. Using the SMC method in the same manner as in Eq.~\eqref{eq:SMC}, the EIG becomes
	\begin{align}
		\mathbb{E}_{\mathtt{Dist}_{\text{KL}}}(\tau, D')=& -\bigg[1-\sum_{k=1}^{n_\text{p}} w_{k} \text{Pr}(0 \vert (\omega_k, B_k), \tau)\bigg] \nonumber \\
		& \times \sum_{k=1}^{n_\text{p}} w_{k} \log[1 -\text{Pr}(0 \vert (\omega_k, B_k), \tau)] \nonumber \\ 
		&-\sum_{k=1}^{n_\text{p}} w_{k} \text{Pr}(0 \vert (\omega_k, B_k), \tau)\nonumber \\
		&\times \sum_{k=1}^{n_\text{p}} w_{k} \log[\text{Pr}(0 \vert (\omega_k, B_k), \tau)],
	\end{align} 
	and we choose $N_\text{bench}=400$ different combinations of $(\omega, B)$ as well as the coherence time $T_2=170\, \mu\text{s}$ to simulate the experiment. The set of possible values of the control parameter $\tau$ is $\mathcal{C} = \{0.51, 0.52, 0.53,\dots, 7\}\mu\text{s}$. In order to reduce the number of measurements before convergence we set $p_0=p_1=1$ and further details are presented in Appendix~\ref{appendix:setup}. After the simulation, we plot the medians of the relative uncertainty $\Delta\omega$ and $\Delta B$ versus $N_{\text{shot}}$ for our BED (orange squares) and the traditional non-adaptive method (blue circles) in Fig.~\ref{fig:acrel}. Based on the benchmark, we find that our BED outperforms the traditional method even with an imperfect likelihood function, showing its feasibility for this scenario.
	
	\section{Accelerated Computation}
	\label{sec:fastcalc}
	Despite the fact that BED can utilize experimental resources efficiently, it remains a challenge to avoid adverse effects from the computational consumption. When applied to the field of quantum sensing in actual laboratory settings, one might find that the increase in computational time can outweigh the time saved by using optimized measurements. Therefore, several possible solutions have been put forward to mitigate this issue, such as applying BED to simple systems with suitable heuristics~\cite{Santagati2019, Joas2021} and resorting to computer clusters~\cite{Gentile2021}. However, in more generalized systems, previous efforts cannot reduce the time cost of a BED effectively~\cite{Dushenko2020, Gentile2021}. Here, we propose that a single consumer-grade GPGPU along with an asynchronous version of our BED can greatly reduce or eliminate the computational overhead for more general systems and demonstrate the feasibility within the model introduced in Sec.~\ref{ssec:nspins}. 
	
	\subsection{GPGPU acceleration}
	\label{ssec:gpu}
	We first give an overview of the computing resources for our BED.  The computational resources required by our BED scale linearly with the number of particles $n_p$ used in the SMC method. For the system discussed in Sec.~\ref{ssec:nspins}, the averaged time spent by the probe during a single-shot measurement is $\Delta t \approx 270\, \mu\text{s}$, excluding initialization and read-out. The likelihood function Eq.~\eqref{eq:nspinpr} is evaluated $n_p|\mathcal{C}|\approx1.95\times10^5$ times per measurement based on our EIG usage and $n_\text{p} = 3200$. These parameters mean a throughput of roughly $7.2\times10^8$ evaluations per second for the likelihood function, which can easily overwhelm a CPU. Fortunately, the performance of a single GPGPU can satisfy such requirements without the burden of setting up a computer cluster. However, existing algorithms utilizing GPGPUs~\cite{qinfer-1_0} only have inefficient implementations. To deal with this problem, we design the algorithms and relevant programs in-house such that both the SMC method and the EIG efficiently run on the GPGPU~\footnote{Source code is available upon reasonable request.}. Vectorization~\cite{Cebrian2014}, usage of mixed-precision arithmetic~\cite{Baboulin2009}, as well as other techniques are adopted to fully accommodate the power of the GPGPU. In addition, the fragmented communication between the GPGPU and other parts of the instruments is minimized to reduce the overhead. As a result, our GPGPU implementation significantly reduces the time cost of the BED without requiring expensive equipment.
	
	To demonstrate these improvements, we retest the experiments shown in Fig.~\ref{fig:c13dist}. For comparison, an eight-core Intel Core i7 CPU running at $4.6\, \text{GHz}$ and an NVIDIA RTX 2080 Ti GPGPU are used, respectively. Numba~\cite{numba} is used for the CPU, while CuPy~\cite{cupy_learningsys2017} provides software access to the GPGPU. 
	The experiments are repeated $10$ times before recording the averaged time spent by both types of hardware. In each repetition, $N_\text{shot} = 18000$ measurements are conducted, and the evolution time of the probe needs at least $N_\text{shot}\Delta t\approx 4.8\,$s. 
	We find that the CPU spends $223.22\,$s per experimental run and GPGPU completes the same computation within merely $4.26\,$s. The CPU can only achieve a throughput of $3.47 \times 10^7$ evaluations per second after ruling out overhead, while the GPGPU can reach $7.18\times 10^9$ evaluations per second in throughput. 
	Based on these results, hundreds of high performance CPU cores are needed to match the performance offered by a single GPGPU, which is consistent with the high performance computer setups in Ref.~\cite{Gentile2021}. This also proves that the practicality of our BED can be significantly improved by our GPGPU implementation.
	
	\subsection{Asynchronous operation}
	\label{ssec:async}
	While GPGPUs can reduce the computation time to levels comparable to the time spent by the probe, the absolute sensitivity of the BED is still vulnerable to computational overhead. Traditionally, instruments need to wait for optimized control parameters before measurements in the BED setups, which decreases the utilization of experimental time. As a result, improvements in the absolute sensitivity of the BED are degraded as reported in Ref.~\cite{Dushenko2020}.
	Here, we propose an asynchronous version of a BED in an effort to hide computation time. In our asynchronous BED setup, the instruments no longer wait for optimal controls inferred from the EIG $\mathbb{E}_{\mathtt{Dist}_{\text{KL}}}(c, D')$. The optimization process starts before the instruments complete their operation for the previous iteration, such that the optimized control parameters will be available upon the beginning of the next iteration. As a compromise, the control parameters of the instruments are based on all but the most recent measurements, i.e., $c_i = \text{argmax}_{c \in \mathcal{C}} \mathbb{E}_{\mathtt{Dist}_{\text{KL}}}(c, D'_\text{old})$. Here, ${D'}_\text{old}=(d_1,\dots, d_{i-1-T})$ signifies that the results of the last $T$ measurements are ignored during the optimization.
	As we will show, this technique has minimal negative impact on the relative uncertainty of the parameter-estimation results.
	
	\begin{figure}[t]
		\centering
		\includegraphics[width=\linewidth]{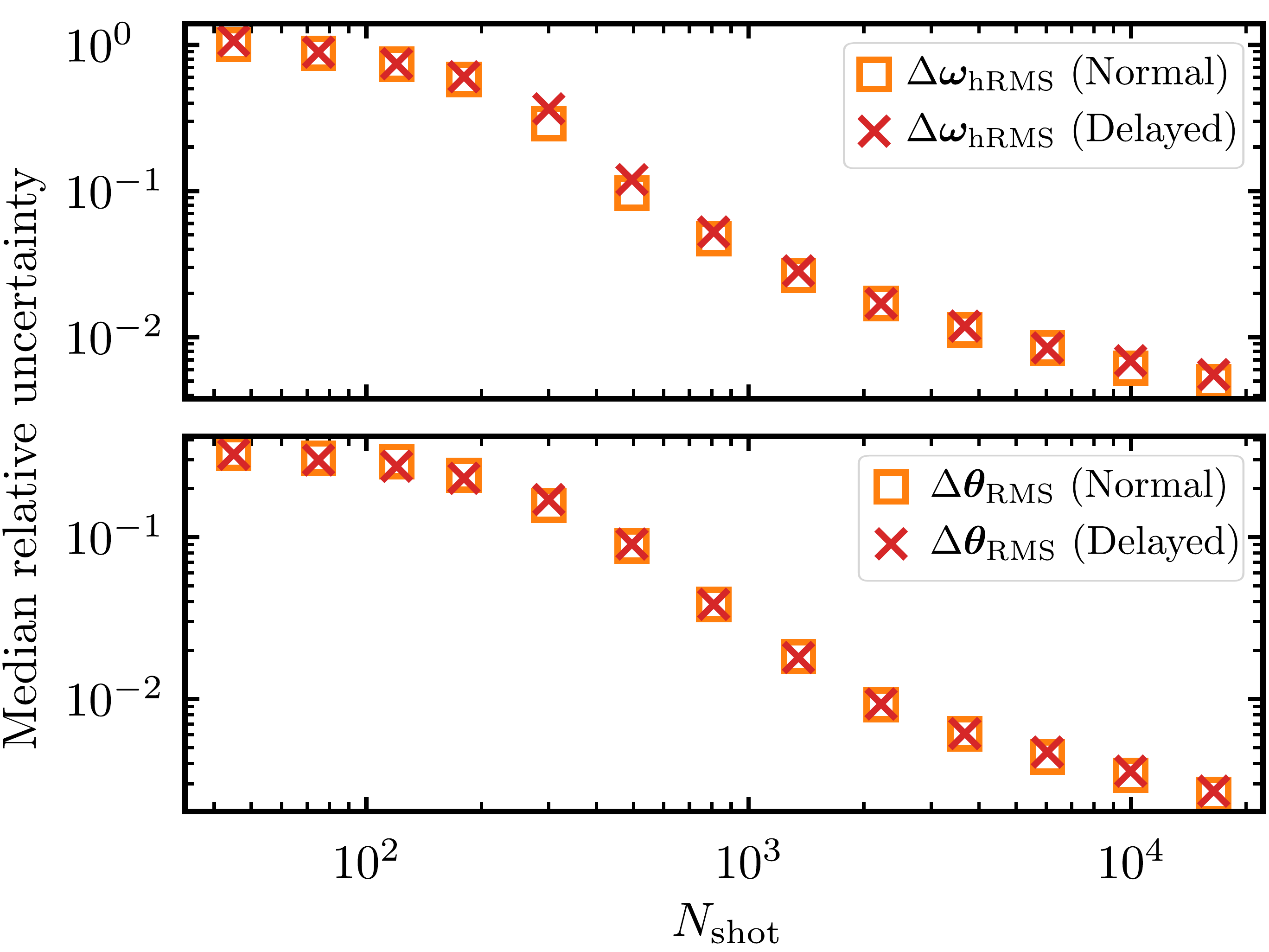}
		\caption{Comparison of the performance between the normal (orange squares) and asynchronous (red crosses) version of the BED for sensing $n_\text{C}=2$ nuclear spins based on $N_\text{bench}=900$ different combinations of $(\boldsymbol{\omega}_\text{h},\boldsymbol{\theta})$. The asynchronous version of the BED has a delay of $T\Delta t\approx4\, \text{ms}$ in terms of the collected measurement results. Top: Median of relative uncertainty $\Delta\boldsymbol{\omega}_\text{hRMS}$ versus the number of single-shot measurements $N_\text{shot}$. Bottom: Median of the relative uncertainty $\Delta\boldsymbol{\theta}_\text{RMS}$ versus $N_\text{shot}$.}
		\label{fig:nspinlatent}
	\end{figure}
	
	To test the degradation introduced by this technique, we rerun the test from Sec.~\ref{ssec:nspins} with $n_\text{C}=2$ and $T=15$ after changing the EIG to $\mathbb{E}_{\mathtt{Dist}_{\text{KL}}}(c, D'_\text{old})$. A sufficiently large delay~\cite{Santagati2019} of $T\Delta t\approx4\,\text{ms}$ is in place. In Fig.~\ref{fig:nspinlatent}, we compare the $\Delta\boldsymbol{\omega}_\text{hRMS}$ and $\Delta\boldsymbol{\theta}_\text{RMS}$ obtained from an asynchronous operation to those from Fig.~\ref{fig:figc13rnd2rel}. It can be observed that the comparison shows only minimal discrepancies and that the degradation caused by our asynchronous operation is negligible. By combining our asynchronous operation with the use of a GPGPU, we can eliminate the overhead of our BED and obtain a reduction of up to $90\%$ in the total time cost, as compared with traditional methods.

	\section{Conclusion}
	\label{sec:sum}
	In this work, we have proposed a practical real-time adaptive quantum-sensing protocol guided by Bayesian experimental design. Instead of heuristics, the model-agnostic EIG was used as the utility function. The performance of this protocol was analyzed for sensing nuclear spins and oscillating magnetic fields using an NV-center probe. Compared with traditional methods, our protocol can significantly reduce the required experimental time and improve the sensitivity of the measurements. Besides, we demonstrated two techniques to reduce the overhead brought by computation and communication delay. We introduce a GPGPU acceleration and an asynchronous operation to further enhance the performance of the Bayesian experimental design and obtain a reduction of up to $90\%$ in total time cost. With this significant reduction, we believe the scheme presented here supplies viable prospects for the improvement of quantum-sensing experiments.
	\section{acknowledgments} 
	This work is supported by the National Natural Science Foundation of China (Grant No.~12161141011, No.~11874024, and No.~11690032), the National Key R$\&$D Program of China (Grant No. 2018YFA0306600), the Fundamental Research Funds for the Central Universities, the Open Project Program of the Shanghai Key Laboratory of Magnetic Resonance, and the Interdisciplinary Program of the Wuhan National High Magnetic Field Center (Grand No. WHMFC202106). D.-X. Li is also funded by the China Postdoctoral Science Foundation (Grant No. 2021M690062).
	
	\appendix
	
	\section{Simulation setup details}
	\label{appendix:setup}
	For all simulations done in this paper, the basic dynamical-decoupling sequence with eight pulse units is given by $R_y(\pi/2)-\tau-R_x(\pi)-2\tau-R_y(\pi)-2\tau-R_x(\pi)-2\tau-R_y(\pi)-2\tau-R_y(\pi)-2\tau-R_x(\pi)-2\tau-R_y(\pi)-2\tau-R_x(\pi)-\tau-R_y(3\pi/2)$, where $R_\kappa(\theta)$ with $\kappa=x,y$ denotes a rotation of the probe state around the $\kappa$ axis of the Bloch sphere about an angle $\theta$, while the probe evolves freely between pulses. The optimized $\tau_i$ are produced in small batches $\text{C}_\text{batch}=(\tau_i, \tau_{i+1},\dots, \tau_{i+n_\text{batch}-1})$. The elements of $\text{C}_\text{batch}$ are sampled from the distribution $\text{Pr}_\text{batch}(\tau)\propto \left[\mathbb{E}_{\mathtt{Dist}_{\text{KL}}}(\tau, D')\right]^p$, in which $p=6$ is determined by  empirical tests. For the simulations in Sec.~\ref{ssec:nspins} we choose $n_\text{batch}=15$ while $n_\text{batch}=30$ is set for the experiments in Sec.~\ref{ssec:acmag}. 
	
	To generate exact values of parameters that we want to estimate in Sec.~\ref{sec:app}, a combination of sweeping and randomization is used. In detail, for the exact values used in Sec.~\ref{ssec:nspins}, the hyperfine couplings $\boldsymbol{\omega}_\text{h}/2\pi$ are generated by sweeping through $19.0-83.4\, \text{kHz}$, and the angles $\boldsymbol{\theta}$ are sampled from a uniform distribution on $[0,2\pi]$. In Sec.~\ref{ssec:acmag}, on the other hand, we first generate $20$ different frequencies $\omega/2\pi$ by sweeping from $111\, \text{kHz}$ to $1.27\, \text{MHz}$ and then we sweep the magnetic-field strength $B$ from $0.031\omega/\gamma$ to $0.169\omega/\gamma$ for each value of $\omega$.
	
	Finally, the technical aspects of our numerical simulation setup work as follows. We use QInfer~\cite{qinfer-1_0} with modifications crucial for high throughput GPGPU operations. 
	The bounds for the estimation parameters are the same for both our BED and the traditional method. In Sec.~\ref{ssec:nspins}, for both methods the upper and lower bounds for the elements of $\boldsymbol{\omega}_\text{h}/2\pi$ are given by $6\, \text{kHz}$ to $265\, \text{kHz}$, respectively, and the prior distribution given to our BED is a uniform distribution for both $\boldsymbol{\omega}_\text{h}$ and $\boldsymbol{\theta}$. In Sec.~\ref{ssec:acmag}, $B\gamma/\omega$ ranges from $0.013$ to $0.177$, while $\omega$ stays valid as long as $79.6\, \text{kHz}\leq \omega/2\pi \leq 1.35\, \text{MHz}$. We note that while our use of the EIG as the utility function is not sensitive to imprecise prior probabilities~\cite{sep-formal-belief}, imprecise priors might need non-trivial treatment for some tasks, such as quantum-sensing tasks with few experimental runs. Discussions on such topics can be found in Ref.~\cite{Go2022}.
	In all simulations, the number of particles used in the SMC method stays constant throughout the experiment to simplify the discussion related to computational resources in Sec.~\ref{ssec:gpu}. For experiments conducted in Sec.~\ref{ssec:nspins}, $n_\text{p}=6400$ is used for $n_\text{C}=3$, as the increase in computation time is tolerable (the computation time increases $31\%$ when compared with the time recorded for $n_\text{C}=2$ and $n_\text{p}=3200$). For other examples and demonstrations, $n_\text{p}=3200$ is used. As for resamplers used in the SMC method, we chose the Liu-West resampler~\cite{Liu2001} with $a=0.98$ and a naive remapping technique to mitigate ambiguities in the particle distributions. We note that a different resampler, such as the one proposed in Ref.~\cite{Granade_2017}, might be more suitable for these models. The data of our simulations are available in the~\cite{SM}.

    For the traditional method being compared with, we use the same SMC method and the same estimator as those used in the BED. We do not use least-squared methods due to their frequent divergence over the exact values (according to our testing, least-squares-based estimations seldom converge to exact values even when the exact population numbers are supplied). This behavior of poor performance can be attributed to the high dynamic-range and the non-convexity of our demonstrations. In addition, the traditional method does not have access to utility function. Instead, a random number generator governs the control parameters used for each single-shot measurement. Doing so can improve consistency in results when averaged shot count per unique control parameter is small, while having the same effects as parameter-scanning once sufficiently large amount of measurements are conducted.
	
	\section{Likelihood function of the probe sensing AC magnetic field}
	\label{appendix:ac}
	A rigorous formalism of the dynamics of the population of a spin probe subject to an external signal and noise can be found in Refs.~\cite{Degen2017, Pang2017}. Here, we provide, for the sake of completeness, an elementary yet self-contained derivation of Eq.~\eqref{eq:acprideal}. Owing to the XY8 dynamical-decoupling pulse sequence, the probe is protected from external noises with low frequencies, and the modulation function becomes
	\begin{equation}
		h(t) = \begin{cases}
			1, &t \in \left[ (4k-1)\tau, (4k+1)\tau\right) \\
			-1, &t \in \left[ (4k+1)\tau, (4k+3)\tau\right).
		\end{cases}
	\end{equation}
	The probe is initialized in the state $|0\rangle$, and if we neglect additional decoherence, ${\hat{H}}_{\text{noise}}(t)$ will effectively not contribute to the probability of the probe preserving $\left|0\right>$. If we denote the state at time $t=16\tau$, i.e., after the pulse sequence, by $\vert\phi\rangle$ we thus find the probability
	\begin{equation}
	    \vert\langle 0\vert \phi \rangle \vert^2=\frac{1}{2}[1+\cos(a\cos(8\omega\tau+\varphi))]
	\end{equation}
	at the read-out, with $a$ given in Eq.~\eqref{eq:phis}.
	
	As mentioned in the main text, no timestamps are recorded for each measurement and the initial phase $\varphi$ must therefore be averaged out. Assuming the intervals between measurements to follow a uniform distribution, 
	the averaged probability can be obtained according to $\overbar{\vert\langle 0\vert \phi \rangle \vert^2}=\int_0^{2\pi}\vert\langle 0\vert \phi \rangle \vert^2 d\varphi/2\pi$ and reads
	\begin{equation}
	    \overbar{|\left<0|\phi\right>|^2} = \frac{1}{2}[1+J_0(a)] 
	\end{equation}
	with the Bessel function $J_0$. 	Equation~\eqref{eq:acprideal} is then obtained by adding an exponential decay toward a probability $1/2$ with the coherence time $T_2$ to the expression above.
	Since in our situation we have $a \ll 1$, we employ the truncated series expansion
	\begin{equation}
		J_0(a)=
		1-\frac{1}{4}a^2+\frac{1}{64} a^4-\frac{1}{2304} a^6+O(a^8)
	\end{equation} 
	of the Bessel function for the sake of numerical efficiency.

	\bibliography{reference}
\end{document}